\documentclass{article}

\usepackage{siunitx}
\usepackage{booktabs}

\usepackage{tabularx}
\usepackage{PRIMEarxiv}
\usepackage{booktabs}
\usepackage{siunitx} 
\usepackage[utf8]{inputenc} 
\usepackage[T1]{fontenc}    
\usepackage{hyperref}       
\usepackage{url}            
\usepackage{booktabs}       
\usepackage{amsfonts}       
\usepackage{nicefrac}       
\usepackage{microtype}      
\usepackage{lipsum}
\usepackage{fancyhdr}       
\usepackage{graphicx}       
\usepackage{hyperref}
\graphicspath{{media/}}     

\usepackage{float}      
\usepackage{tabularx}    
\usepackage{booktabs}
\usepackage{tabularx} 
\usepackage{booktabs} 
\usepackage{times}
\usepackage{soul}
\usepackage{url}
\usepackage[small]{caption}
\usepackage{graphicx}
\usepackage{amsmath}
\usepackage{subcaption}
\usepackage{amsthm}
\usepackage{booktabs}
\usepackage{algorithm}
\usepackage{algorithmic}
\usepackage[switch]{lineno}
\usepackage{multirow} 
\usepackage{xcolor}   
\usepackage{amssymb}

\title{Ti-Audio: The First Multi-Dialectal End-to-End Speech LLM for Tibetan
}

\author{
  \textbf{Jialing Wang$^{1,2}$, Yue Zhao$^{1, \star}$, Yuhao Zhang$^{2}$} \\
  \textbf{Jing Yu$^{1}$, Shaosai Li$^{1}$, Zhanchen Dai$^{2}$, Benyou Wang$^{2}$, Haizhou Li$^{2}$} \\
  $^{1}$Minzu University of China, Beijing, China \\
  $^{2}$The Chinese University of Hong Kong, Shenzhen, China \\
  \texttt{zhaoyueso@muc.edu.cn} \\
}

\begin{document}
\maketitle

\begin{abstract}
Recent advances in Speech Large Language Models (Speech-LLMs) have made significant progress, greatly enhancing multimodal interaction capabilities.However, their application in low-resource and dialect-diverse environments still faces challenges. The severe scarcity of Tibetan data, coupled with the phonetic differences among its major dialects (Ü-Tsang, Amdo, and Kham), is a prime example of this challenge. This paper proposes Ti-Audio, the first multi-dialectal end-to-end  Speech-LLM for Tibetan. To efficiently align speech and text, we introduce a Dynamic Q-Former Adapter that extracts essential acoustic features from variable-length speech, ensuring stable cross-modal alignment even with limited data. At the data level, we leverage mutual assistance among related dialects to alleviate data scarcity and employ a temperature-based sampling strategy to maximize this synergy.  Experimental results demonstrate that Ti-Audio achieves state-of-the-art performance on Tibetan benchmarks for automatic speech recognition and speech translation.  Our work validates the effectiveness of cross-dialectal cooperation and provides a scalable paradigm for the development of Speech-LLM in low-resource scenarios.
\end{abstract}

\section{Introduction}
As the capabilities of Large Language Models (LLMs) continue to expand \cite{brown2020language}, research focus has shifted from text-only understanding to models capable of handling multiple modalities \cite{alayrac2022flamingo,openai2024gpt4o}. Since speech is the most fundamental form of human interaction, Speech-LLMs have become a key branch of multimodal large models, primarily by combining advanced speech encoders with LLM \cite{zhang2023speechgpt,chu2024qwenaudio}. However, the construction of existing Speech-LLMs heavily relies on high-resource corpora \cite{tang2024salmonn,radford2023robust}. In contrast,exploration of systematic approaches for low-resource languages and specific dialects remains severely insufficient \cite{pratap2024scaling}, leaving a critical gap in the era of generative AI.

Developing Speech-LLMs in low-resource scenarios faces two major challenges. 
First, in terms of architecture, existing speech models are typically developed in high-resource languages, where data-driven training strategies rely on massive datasets to ensure stable modality alignment and generalization capabilities \cite{zhang2023speechgpt,tang2024salmonn,pratap2024scaling}. 
However, in low-resource environments, directly adopting high-resource architectures often fails to achieve the desired results \cite{radford2023robust,huang2024investigating}. Therefore, efficient training architectures for low-resource Speech-LLMs remain largely unexplored. Second, at the data level, multilingual processing in the text domain has demonstrated that related languages within the same family can promote mutual assistance and cross-linguistic transfer \cite{liu2020multilingual}. Accordingly, we investigate whether constructing multi-dialect models within the same language family can effectively validate and leverage similar collaborative mechanisms in the speech domain to alleviate data scarcity.

\begin{figure}[t]
   \centering
   \includegraphics[width=0.5\linewidth]{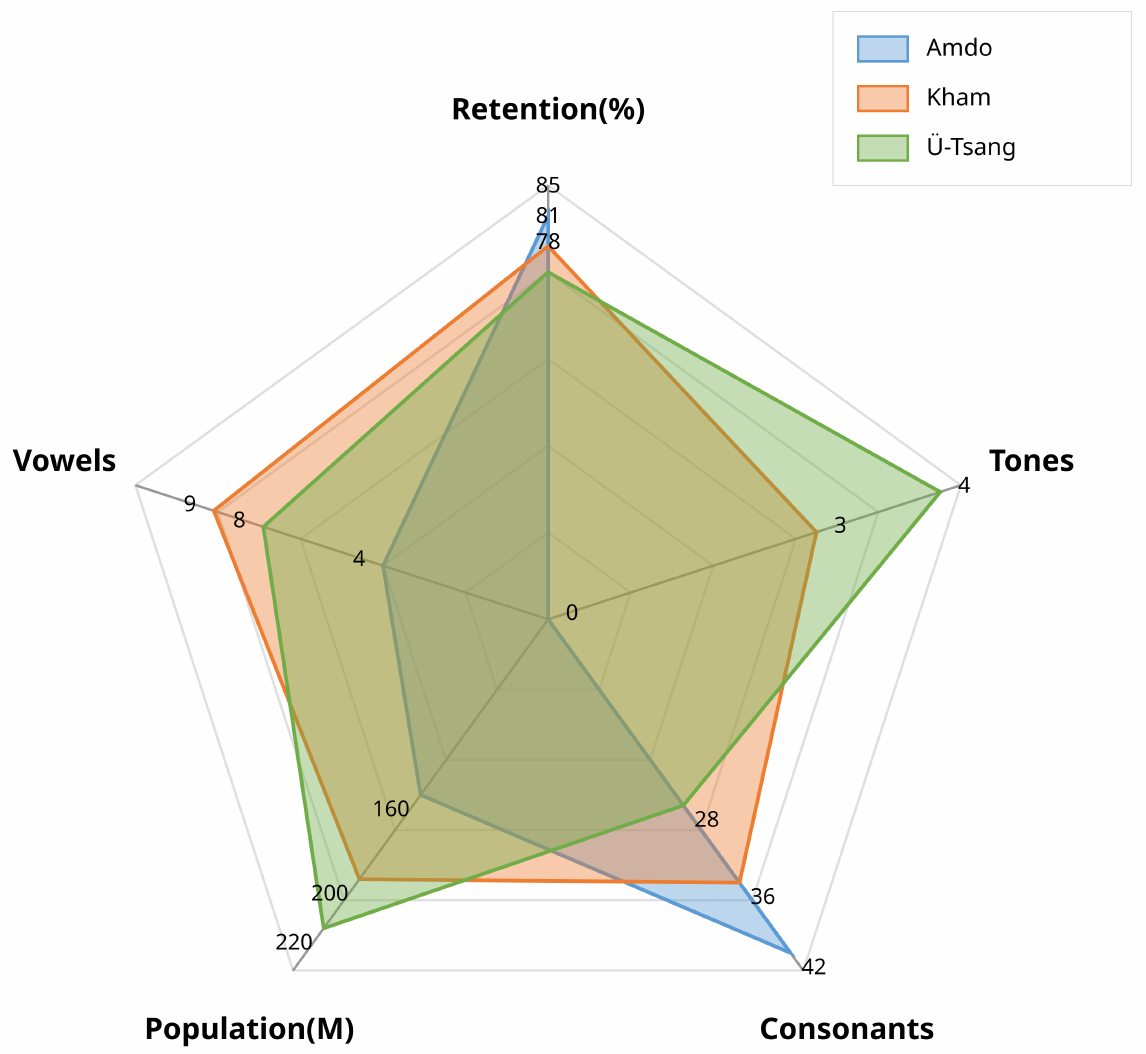}
   \caption{Linguistic divergence among Tibetan dialects}
   \label{fig:dialect_radar}
\end{figure}
We chose Tibetan as a representative case study to conduct our research. 
On the one hand, Tibetan is a typical low-resource language, with the scale of publicly available annotated speech data limited to approximately 600 hours \cite{li2020msr,duan2021tibmd}. 
On the other hand, Tibetan exhibits significant dialectal differences. Its major dialects—Ü-Tsang, Amdo, and Kham—demonstrate significant differentiation across multiple linguistic dimensions, including tonal systems and phonological structures, as shown in Figure \ref{fig:dialect_radar} \cite{tournadre2003manual}.

Inspired by the Q-Former method in low-resource vision tasks \cite{li2023blip2} and the dynamic length characteristics of speech sequences, we design a Dynamic Q-Former Adapter for speech processing. This adapter can effectively extract key acoustic information while compressing long sequences, thus ensuring stable alignment of the model in low-resource environments. Secondly, at the data level, we build a multi-dialect speech-LLM model to examine whether imilar cross-dialect cooperative mechanism exists and whether this mechanism can alleviate the data scarcity problem in low-resource environments. In addition, we incorporate a temperature-based sampling strategy to alleviate data imbalance between different dialects, thereby maximizing the cooperative benefits and improving overall performance.

Extensive experiments demonstrate that Ti-Audio achieves state-of-the-art performance in Tibetan speech tasks. Specifically, the model reaches an average BLEU of 22.05 in Speech Translation (ST) and reduces the word error rate (WER) by more than 12\% in automatic speech recognition (ASR). Furthermore, end-to-end multi-dialect training yields consistent performance improvements across all Tibetan dialects, validating a stable cooperative mechanism in the speech domain. Comparative analysis also show that our dynamic Q-Former adapter provides significantly higher performance improvements than standard linear layers, highlighting its crucial role in efficient, low-resource cross-modal modeling.

Our main contributions are summarized as follows:
\begin{itemize}
    \item We propose Ti-Audio, the first \textbf{end-to-end Tibetan Speech-LLM}. Compared to existing cascaded systems, Ti-Audio achieves \textbf{state-of-the-art} performance in both Tibetan ASR and ST.
    
    \item We design a \textbf{Dynamic Q-Former}, an adapter specifically designed to efficient bridge the gap between the speech encoder and the LLM under resource-constrained conditions.
    
    \item We experimentally validate \textbf{cross-dialect positive transfer} between relevant Tibetan dialects under unified training conditions. We further employ a temperature-based sampling scheme to better utilize this cross-dialect positive transfer.
    
\end{itemize}


\section{Related Work}
\subsection{Speech-LLMs} 
Early LLMs for speech understanding primarily relied on cascaded pipelines, which combine an ASR module with a text-based LLM \cite{huang2024audiogpt,koneru2024blendingllmscascadedspeech}. While these methods effectively utilize mature ASR and LLM techniques, they suffer from error propagation problems and fail to fully leverage paralinguistic information, such as speaker features or emotional expressions\cite{bahar2020tight,min2025when}.
To overcome these limitations, recent studies have shifted towards E2E architectures, which typically integrate pre-trained speech encoders such as wav2vec and HuBERT \cite{NEURIPS2020_92d1e1eb,hsu2021hubert} to directly map speech signals to the semantic space of LLMs. For example, AudioPaLM and SpeechGPT leverage discrete audio tokens to enable cross-modal conversation within unified transformers \cite{rubenstein2024audiopalm,zhang2023speechgpt}. Building upon this paradigm, subsequent works such as Video-LLaMA, SALMONN and Qwen-Audio introduce learned adapters, e.g., Q-former to selectively extract informative acoustic features and enhance cross-modal alignment, thereby improving generalization across diverse speech understanding tasks \cite{zhang-etal-2023-video,tang2024salmonn,chu2024qwenaudio}.

\subsection{Low-Resource Language Modeling}
Despite robustness, current E2E Speech-LLMs significantly depend on large-scale speech corpora pre-training, often failing to generalize to low-resource languages \cite{fong2025speech,cui-etal-2025-recent}. To tackle this, SoundWave proposes a dual-adapter architecture, achieving SOTA performance on English while using only 2\% of the training data required by Qwen2-Audio \cite{chu2024qwen2audio,zhang2025soundwave}. Similarly, XLSR-Thai explores external alignment of encoder and adapter to avoid costly back-propagation of the full LLM during training \cite{shao2025towards}. However, these approaches primarily focus on language-specific learning, where each language is learned independently from its own corpus. Consequently, the potential mutual information across phonetically diverse linguistic variants has been largely overlooked, creating a significant gap in cross-dialectal speech modeling \cite{zhou-etal-2024-dialectmoe}.

Beyond language-specific tuning, efficient modality alignment is critical for model initialization in data-constrained scenarios. In the field of vision, the Q-former in BLIP-2 \cite{li2023blip2} proven to effectively extract salient features in low-resource settings. Meanwhile, recent research \cite{zhou2024dapt} introduced a dynamic that adaptively adjusts parameters based on different input conditions, providing a lightweight mechanism scalable to multimodal scenarios. However, these architectures are designed for static visual signals and cannot be directly applied to speech with dynamic temporal characteristics.

\section{Methodology}

\subsection{Model Architecture}
\begin{figure*}[t]
   \centering
   \includegraphics[width=0.9\textwidth]{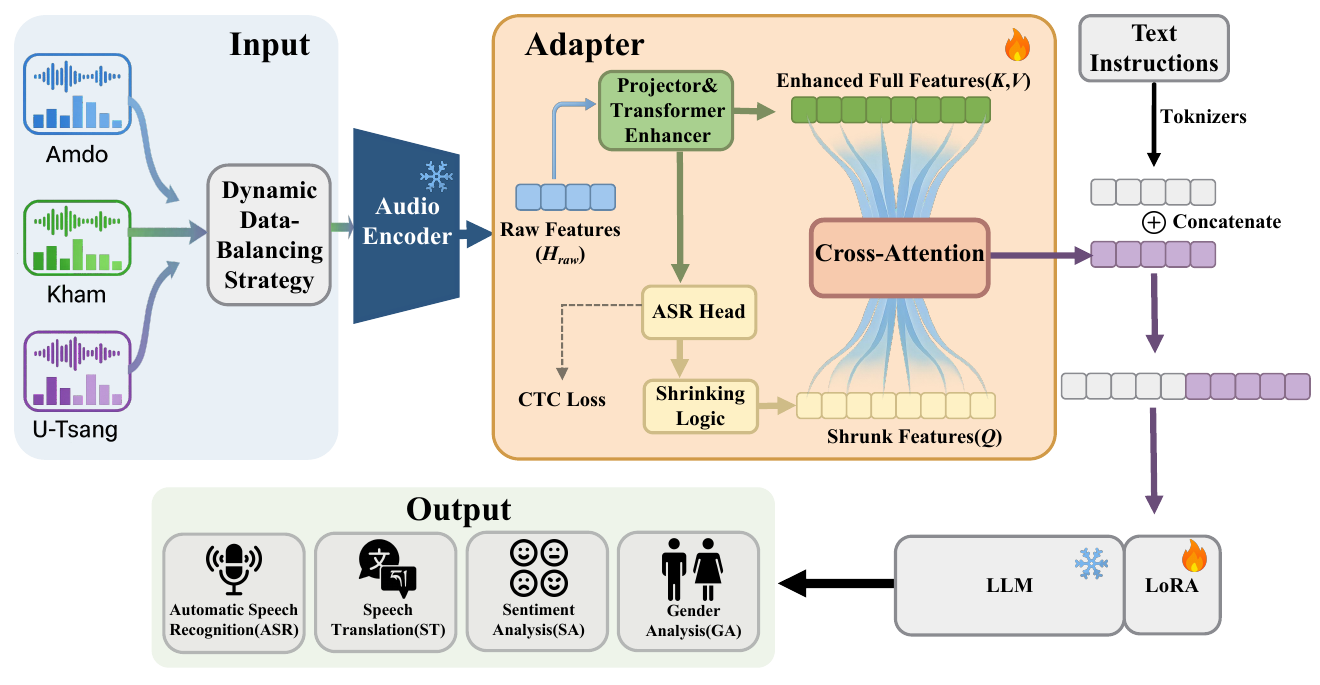}
   \caption{The model structure of Ti-Audio}
   \label{fig:model_architecture}
\end{figure*}

The overall architecture of Ti-Audio is shown in Figure~\ref{fig:model_architecture}. This framework consists of three parts: Multi-Dialect Perception, Dynamic Q-Former Adapter and Generative Reasoning.

\subsubsection{Multi-Dialect Perception}
The framework accepts raw waveforms from diverse Tibetan dialects (Ü-Tsang, Amdo, and Kham). To maximize the dialect mutual assistance mechanism, we employ a Temperature-Aware Data Balancing Strategy before feeding inputs into a frozen Audio Encoder to extract continuous acoustic features $H_{raw}$.

\subsubsection{Dynamic Q-Former Adapter}
We propose the \textbf{Dynamic Q-Former Adapter} to address the low-resource alignment problem. Unlike the fixed resolution of the image domain, speech signals are inherently dynamic, necessitating the design of an adapter that can dynamically adapt to changes in speech. Our module bridges this gap by generating dynamic queries directly from the audio input. This approach compresses redundant $H_{raw}$ into information-dense semantic tokens. Therefore, it effectively decouples representation alignment from length alignment, ensuring robust performance in resource-limited settings.

\subsubsection{Generative Reasoning} 
The aligned speech embeddings are concatenated with tokenized text instructions. This unified sequence is then fed into a LLM fine-tuned with Low-Rank Adaptation (LoRA) \cite{hu2022lora} to generate an end-to-end target response.

\subsection{Multi-Dialect Perception}
Tibetan dialect speech faces a severe data imbalance problem, mainly manifested in two aspects: dialect differences, such as the difference between mainstream dialects like Ü-Tsang and rare dialects like Kham, and task distribution, such as the abundance of ASR data and the limited amount of paralinguistic samples. To prevent the model from overfitting to the mainstream category and to maximize the dialect assistance mechanism, we employ a Temperature-Aware Data Balancing Strategy at the input stage. We define a data category $c$ as a unique combination of dialect and task, for example, Kham\_ASR. To make the training distribution more uniform, the sampling probability $p_c$ is formulated as:\begin{equation}p_c \propto \left( \frac{1}{N_c} \right)^{\frac{1}{\tau}}\end{equation}where $N_c$ is the number of samples and $\tau$ is the temperature hyperparameter. Setting $\tau > 1$ effectively upsamples rare dialect-task pairs. This fosters a synergistic effect, enabling the model to leverage general acoustic patterns learned from high-resource tasks to enhance performance in resource-constrained scenarios. Finally, the balanced inputs are processed by a frozen Audio Encoder to extract continuous acoustic representations ($H_{raw}$).

\subsection{The Dynamic Q-Former Adapter} \label{sec:adapter} The standard alignment mechanism, such as Q-Former, relies on a large amount of paired data to learn implicit cross-modal mappings, which can lead to overfitting in data-scarce scenarios. To address this, we propose the Dynamic Q-Former Adapter. As shown in Figure~\ref{fig:model_architecture}, this module uses a Dynamic Q-Former to extract key semantic frames from the raw audio, transforming continuous acoustic features ($H_{raw}$) into LLM-understandable tokens. This strategy significantly improves data utilization efficiency and ensures robust alignment even in data-scarce environments.

\subsubsection{Context Enhancement ($K, V$)}
As shown in the green module of Figure~\ref{fig:model_architecture}, we employ a \textbf{Projector and Transformer Enhancer} to process the raw acoustic features $H_{raw}$, thereby capturing high-level semantic context. Crucially, this enhanced representation serves a dual purpose. First, it is preserved as the full sequence of Keys ($K$) and Values ($V$) to maintain fine-grained acoustic integrity, capturing key prosodic cues such as tone that are typically lost during compression. Simultaneously, these features are forwarded to an auxiliary ASR head to compute frame-level speech probabilities, thus providing the necessary structural guidance for subsequent dynamic query generation.

\subsubsection{Dynamic Query Generation ($Q$)}
Standard Q-Former relies on a fixed number of static query vectors, making it difficult to adapt to variable-length speech and prone to overfitting in resource-constrained scenarios. To address this, we propose Dynamic Q-Former, which generates instance-specific queries through an adaptive mechanism.

Specifically, the enhanced features ($K, V$) are projected through an auxiliary ASR head. The output probabilities also serve a dual purpose: computing the CTC loss for supervision and acting as structural signals for our Shrinking Logic. We leverage the inherent ''peak'' characteristic of CTC—that effective phonemes are represented as probability peaks—to accurately locate semantic boundaries. This logic preserves these high-information frames while filtering out redundant whitespace; the resulting sequence constitutes our dynamic query ($Q$). By transforming query acquisition from learning parameters from scratch to an explicit selection guided by ASR, this approach introduces a strong inductive bias, significantly reducing the reliance on massive training data.

\subsubsection{Cross-Modal Fusion} 
Finally, we employ a Cross-Attention mechanism to fuse the generated components. The compact Dynamic Queries ($Q$) serve as semantic anchors that ``look back'' at the preserved full-context Keys and Values ($K, V$) from Step 1 to retrieve missing details:
\begin{equation}
Z = \text{Attention}(Q, K, V) = \text{softmax}\left(\frac{QK^T}{\sqrt{d_k}}\right)V
\end{equation}
This mechanism ensures that the final representation $Z$ is both compact and informative, providing a powerful solution for low-resource alignment.

\subsection{Training Strategy}
To jointly optimize the generation capability of the LLM and the alignment accuracy of the adapter, we employ a multi-task learning objective. The total loss $\mathcal{L}$ is formulated as follows:
\begin{equation}\mathcal{L} = \mathcal{L}_{NTP} + \lambda \times \mathcal{L}_{CTC}\end{equation}
The primary loss term $\mathcal{L}_{NTP}$ represents the standard next-word prediction loss. It utilizes cross-entropy via LoRA to optimize the LLM parameters and adapter projection, ensuring that the model generates coherent and context-accurate text responses under aligned speech input conditions.

The second loss term $\mathcal{L}_{CTC}$ is the connection-temporal classification loss applied to the auxiliary ASR head in the selection path. This term plays a structural role in our framework. It supervises the ASR head, enabling it to learn accurate speech alignment from the raw features.

\section{Experiments}

This section evaluates Ti-Audio. We first present the experimental setup and data pipeline. We then analyze performance on semantic tasks (ASR, ST) and non-semantic benchmarks.

\subsection{Experimental Setup}

\subsubsection{Model Settings}

In the training, the audio encoder is mHuBERT-147 \cite{zanonboito2024mhubert}, and the foundation model is TiLamb-7B \cite{wenhao2024tilamb}, derived from LLaMA2 \cite{touvron2023llama}. The Dynamic Adapter consisted of a Projector and Transformer Enhancer (mapping 768 to 4096 dimensions with 32 heads and an 8192 FFN dimension), an ASR Head facilitating the Shrinking Logic, and a Cross-Attention module for detail retrieval. We apply LoRA \cite{hu2022lora} to the Attention modules ($q, k, v$), where the rank $r=64$ and $\alpha=16$, while fine-tuning all parameters of the adapter. For semantic alignment, the CTC head is initialized with TiLamb text embeddings with a learnable logit scale (initialized to 2.3026) and CTC loss weights $w=0.3$. To address the dialect imbalance problem and maximize the dialect cross-assistance mechanism, a dynamic data balancing strategy with a sampling temperature of $T=3.0$ is employed. The experiments are conducted on NVIDIA H200 GPUs using BF16 precision and Flash Attention \cite{NEURIPS2020_92d1e1eb}. The model is optimized using AdamW optimizer \cite{loshchilov2017decoupled} with a learning rate of $1.4 \times 10^{-4}$ and a cosine scheduler, with a batch size of 64 per device.

In the evaluation, Ti-Audio is evaluated on Tibetan ASR and ST, speaker emotion recognition(SER), and speaker gender recognition(GR) tasks across the Amdo, Kham, and Ü-Tsang dialects. To ensure data integrity, repeated samples are removed before training to avoid data leakage. We compare Ti-Audio with SOTA cascaded systems (mHuBERT + LLM) and massive multilingual models such as Meta MMS \cite{pratap2024scaling} to demonstrate its superior cross-dialectal generalization and semantic rectification capabilities.

\subsubsection{Dataset Construction and Statistics}
Recent studies rely heavily on proprietary data. Consequently, standardized public benchmarks are lacking in this field. To address this challenge, we developed a rigorously selected dataset. The MUC-Tibetan-Speech-LLM dataset\footnote{\url{https://github.com/zi123l/MUC-Tibetan-Speech-LLM-Test-Dataset}\label{fn:github}} dataset contains proprietary data from our lab. Table \ref{tab:datasets} lists comprehensive statistics. The dataset structure ensures comprehensive data coverage. It primarily contains ASR data (496.06 hours) and ST data (377.95 hours). Additionally, it includes supplementary data for GR (75.03 hours) and SER (2.72 hours), with a focus on paralinguistic attributes.

\begin{table}[t]
    \centering
    \small 
    \renewcommand{\arraystretch}{1.1} 
    \setlength{\tabcolsep}{3pt} 

    \caption{Statistics of Tibetan datasets used in this work.}
    \label{tab:datasets}

    \begin{tabularx}{0.6\textwidth}{l X rr}
        \toprule
        \textbf{Task} & \textbf{Data Source} & \textbf{Hours} & \textbf{Samples} \\
        \midrule
        ASR / ST      & MUC-Tibetan-Speech-LLM dataset\footref{fn:github}             & 305.7           & 162k+ \\
        ASR / ST / GR & M2ASR \cite{li2017free}            & \phantom{0}72.3 & \phantom{0}58k+ \\
        ASR           & TIBMD@MUC \cite{TIBMD-MUC}         & \phantom{0}84.3 & \phantom{0}68k+ \\
        ASR           & XBMU-AMDO31 \cite{zhao2021xbmu}     & \phantom{0}31.0 & \phantom{0}22k+ \\
        ASR           & MUC\_greeting \cite{MUC-Greeting}  & \phantom{00}2.8 & \phantom{00}3k+ \\
        ASR / SER     & Tibetan SER \cite{chen2024emotion} & \phantom{00}2.7 & \phantom{00}6k+ \\
        \midrule
        \textbf{Total} &                                   & \textbf{498.8}  & \textbf{319k+} \\
        \bottomrule
    \end{tabularx}
\end{table}

\begin{table}[h]
    \centering
    \small 
    \renewcommand{\arraystretch}{1.1} 
    \setlength{\tabcolsep}{3.2pt} 

    \caption{Dialect distribution statistics.}
    \label{tab:dialect_stats}

    \begin{tabular}{l rrr rrr}
        \toprule
        & \multicolumn{3}{c}{\textbf{Training Set}} 
        & \multicolumn{3}{c}{\textbf{Test Set(ASR/ST)}} \\
        \cmidrule(lr){2-4} \cmidrule(lr){5-7} 
        \textbf{Dialect} 
        & \textbf{Hours} & \textbf{Samples} & \textbf{Ratio} 
        & \textbf{Hours} & \textbf{Samples} & \textbf{Ratio} \\
        \midrule
        Amdo    & 124.4 & 80k & 26.9\% & 8.7  & 5.6k & 23.9\% \\
        Ü-Tsang & 216.9 & 139.5k & 46.9\% & 13.9 & 8.9k & 38.0\% \\
        Kham    & 115.1 & 74.1k  & 24.9\% & 13.9 & 9.0k & 38.1\% \\
        Unknown & 5.9   & 3.9k   & 1.3\%  & --   & --   & --     \\
        \midrule
        \textbf{Total} 
        & \textbf{462.3} & \textbf{297.5k} & \textbf{100\%} 
        & \textbf{36.5} & \textbf{23.5k} & \textbf{100\%} \\
        \bottomrule
    \end{tabular}
\end{table}

We constructed a balanced training corpus. As shown in Table \ref{tab:dialect_stats}, the ratio of Amdo, Ü-Tsang, and Kham languages in this corpus is approximately 1:1.79:1. Concurrently, the test set was constructed with strict uniformity to ensure the fairness of cross-dialect evaluation.

\subsection{Main Results}
We compared Ti-Audio against cascaded systems (mHuBERT + LLM) and dedicated acoustic models. Table \ref{tab:main_results} summarizes its performance across ST, MT, and ASR tasks.

\subsubsection{End-to-End Superiority in Speech Translation}
In the speech translation (ST) task, Ti-Audio outperformed all baseline models and achieved a peak score of 22.05. Crucially, its performance surpasses the cascaded counterparts. This result underscores the effectiveness of the unified end-to-end architecture, indicating that this framework effectively mitigates error propagation inherent in hierarchical systems. Compared with the general large model Gemini 3 Flash (audio input model), Ti-Audio significantly outperforms it. This result verifies the robustness of Ti-Audio in low-resource scenarios.

In the pure machine translation (MT) task, this model outperforms most MT baselines, such as DeepSeek V3.1 and Hunyuan-MT-7B. Although it remains below the 25.99 score of Gemini 3 Flash (MT version), it still demonstrates that Ti-Audio can effectively bridge the cross-modal gap.

\subsubsection{Cross-Dialect Transfer }
Notably, the resource-scarce Kham dialect performed only 1.34 points worse than the dominant Ü-Tsang dialect, with BLEU scores of 22.11 and 23.45 respectively. This empirical finding provides strong support for the language bridge hypothesis, demonstrating that the transitional phonological features of Kham can be effectively utilized. Robust cross-dialect knowledge transfer is achieved through a dialectal mutual assistance mechanism centered on kham, thus mitigating the challenges posed by limited data availability.

\subsubsection{Semantic Correction for ASR}
Ti-Audio demonstrates superior performance in Automatic Speech Recognition (ASR) compared to traditional acoustic baseline models. The finely tuned mHuBERT (CTC) benchmark achieves a word error rate (WER) of 26.77\%, while Ti-Audio's WER reaches 14.46\%, a reduction of over 12\%. This performance improvement highlights the semantic error correction capabilities inherent in the core architecture of Large Language Models (LLMs). Unlike context-independent acoustic models that struggle to eliminate homonyms, Ti-Audio leverages its inherent linguistic prior knowledge to resolve speech ambiguities. Furthermore, Ti-Audio significantly outperforms the large-scale multilingual benchmark Meta Omnilingual (WER 73.04\%). These results further validate the model's ability to handle cross-dialect transfer.

\begin{table*}[t]
    \centering
    \footnotesize 
    \renewcommand{\arraystretch}{1.15}
    \setlength{\tabcolsep}{2.5pt} 

    \caption{Performance comparison on ST, MT, and ASR.}
    \label{tab:main_results}

    \begin{tabularx}{\textwidth}{@{} l X *{4}{S[table-format=2.2]} c @{}}
        \toprule
        \textbf{Task} & \textbf{Model} & \textbf{Amdo} & \textbf{Kham} & \textbf{Ü-Tsang} & \textbf{Avg} & \textbf{Eval Metrics} \\
        \midrule
        \multirow{6}{*}{\textbf{ST}} 
            & Cascaded mHuBERT--DeepSeek V3.1  & 10.02 & 11.69 & 11.71 & 11.14 & BLEU ($\uparrow$) \\
            & Cascaded mHuBERT--Hunyuan-MT-7B  & 9.97  & 11.28 & 11.79 & 11.01 & BLEU ($\uparrow$) \\
            & Cascaded mHuBERT--Gemini 3 Flash & 20.80 & 21.69 & 21.49 & 21.32 & BLEU ($\uparrow$) \\
            & Cascaded mHuBERT--Monlam         & 11.00 & 10.88 & 9.41  & 10.43 & BLEU ($\uparrow$) \\
            \cmidrule(lr){2-7}
            & Gemini 3 Flash           & 1.64  & 2.52  & 4.27  & 2.81  & BLEU ($\uparrow$) \\
            & \textbf{Ti-Audio}                & \textbf{20.59} & \textbf{22.11} & \textbf{23.45} & \textbf{22.05} & BLEU ($\uparrow$) \\
        \midrule
        \multirow{5}{*}{\textbf{MT}} 
            & DeepSeek V3.1          & 14.45 & 15.85 & 15.93 & 15.41 & BLEU ($\uparrow$) \\
            & Hunyuan MT 7B          & 13.23 & 12.51 & 13.83 & 13.19 & BLEU ($\uparrow$) \\
            & Gemini 3 Flash  & 25.03 & 26.53 & 26.42 & 25.99 & BLEU ($\uparrow$) \\
            & Monlam                 & 15.66 & 15.58 & 13.23 & 14.82 & BLEU ($\uparrow$) \\
        \midrule
        \multirow{3}{*}{\textbf{ASR}} 
            & mHuBERT (CTC)    & \multicolumn{1}{c}{27.26 / 10.20} & \multicolumn{1}{c}{26.32 / 10.38} & \multicolumn{1}{c}{26.72 / 11.78} & \multicolumn{1}{c}{26.77 / 10.79} & WER / CER ($\downarrow$) \\
            & Meta Omnilingual & \multicolumn{1}{c}{76.89 / 47.73} & \multicolumn{1}{c}{75.60 / 49.44} & \multicolumn{1}{c}{66.63 / 41.24} & \multicolumn{1}{c}{73.04 / 46.14} & WER / CER ($\downarrow$) \\
            & \textbf{Ti-Audio} & \multicolumn{1}{c}{\textbf{14.25 / 9.57}} & \multicolumn{1}{c}{\textbf{14.15 / 9.60}} & \multicolumn{1}{c}{\textbf{14.99 / 10.31}} & \multicolumn{1}{c}{\textbf{14.46 / 9.83}} & WER / CER ($\downarrow$) \\
        \bottomrule
    \end{tabularx}
\end{table*}

\subsection{Speech Subtask}
We evaluated the speaker emotion recognition (SER) and gender recognition (GR) paralinguistic tasks using limited training data.
\subsubsection{Gender Recognition (GR)}

\vspace{0.5em}

\begin{minipage}[t]{0.48\textwidth}
\vspace{0pt}

\raggedright 

The results shown in Table \ref{tab:gr_performance} indicate that Ti-Audio achieved an accuracy rate of 100.00\% on the female samples. The performance gap of Ti-Audio on the male samples compared to Qwen3-omni-flash was only 1.08\%. Overall, Ti-Audio achieved the highest accuracy rate of 99.60\% throughout the benchmark test.

\subsubsection{Speaker Emotion Recognition (SER)}
Table \ref{tab:ser_performance} presents the results of the benchmark test. 
The average recall rate of Ti-Audio is 22.33\%. 
It performed exceptionally well in terms of high arousal emotions, such as the "anger" category which achieved the highest F1 score 41.67\%. 
This indicates that it can effectively model the pitch features (such as pitch and energy).
However, for subtle emotions (such as "happiness"), modeling remains challenging. 
This indicates that fine-grained emotional representation requires explicit supervision.

\centering
    \footnotesize
    \renewcommand{\arraystretch}{1.15}
    \setlength{\tabcolsep}{4pt}

    \captionof{table}{Gender Recognition (GR) Performance Comparison.}
    \label{tab:gr_performance}

    \begin{tabular}{l S[table-format=3.2] S[table-format=3.2] S[table-format=3.2]}
        \toprule
        \textbf{Model} & \textbf{Prec. (\%)} & \textbf{Female (\%)} & \textbf{Male (\%)} \\
        \midrule
        \textbf{Ti-Audio} & \textbf{99.60} & \textbf{100.00} & 98.92 \\
        Gemini 3 Flash  & 77.63 & 66.65 & 90.35 \\
        Qwen3-omni-flash & 99.21 & 99.28 & \textbf{100.00} \\
        \bottomrule
    \end{tabular}

\end{minipage}
\hfill
\begin{minipage}[t]{0.48\textwidth}
\vspace{0pt}
\centering
\footnotesize 
\renewcommand{\arraystretch}{1.15} 
\setlength{\tabcolsep}{2.8pt} 

\captionof{table}{Speaker Emotion Recognition (SER) Performance.}
\label{tab:ser_performance}

\begin{tabularx}{\linewidth}{@{} l X *{3}{S[table-format=2.2]} @{}}
    \toprule
    \textbf{Cat.} & \textbf{Model} & \textbf{Prec.} & \textbf{Rec.} & \textbf{F1} \\
    \midrule
    \multirow{3}{*}{\textbf{Anger}}
     & Ti-Audio          & \textbf{41.67} & \textbf{41.67} & \textbf{41.67} \\
     & Gemini 3 Flash    & 39.86 & 45.83 & 42.64 \\
     & Qwen3-omni-flash  & 46.81 & 18.33 & 26.35 \\
    \cmidrule(lr){1-5}
    \multirow{3}{*}{\textbf{Fear}}
     & Ti-Audio          & \textbf{23.08} & \textbf{37.50} & \textbf{28.57} \\
     & Gemini 3 Flash    & 41.46 & 14.17 & 21.12 \\
     & Qwen3-omni-flash  & 27.08 & 10.83 & 15.48 \\
    \cmidrule(lr){1-5}
    \multirow{3}{*}{\textbf{Happiness}}
     & Ti-Audio          & 18.06 & \textbf{10.83} & \textbf{13.54} \\
     & Gemini 3 Flash    & 20.47 & 21.67 & 21.05 \\
     & Qwen3-omni-flash  & \textbf{24.14} & 5.83 & 9.40 \\
    \cmidrule(lr){1-5}
    \multirow{3}{*}{\textbf{Neutral}}
     & Ti-Audio          & \textbf{40.00} & 10.00 & 16.00 \\
     & Gemini 3 Flash    & 28.57 & 70.00 & \textbf{40.58} \\
     & Qwen3-omni-flash  & 21.19 & \textbf{83.33} & 33.78 \\
    \cmidrule(lr){1-5}
    \multirow{3}{*}{\textbf{Sadness}}
     & Ti-Audio          & 25.93 & \textbf{11.67} & \textbf{16.09} \\
     & Gemini 3 Flash    & 0.00 & 0.00 & 0.00 \\
     & Qwen3-omni-flash  & \textbf{50.00} & 1.67 & 3.23 \\
    \midrule
    \multirow{3}{*}{\textit{Avg.}}
     & Ti-Audio          & 29.75 & 22.33 & 23.17 \\
     & Gemini 3 Flash    & 26.07 & \textbf{30.33} & \textbf{25.08} \\
     & Qwen3-omni-flash  & \textbf{33.84} & 24.00 & 17.65 \\
    \bottomrule
\end{tabularx}
\end{minipage}

\section{Analysis}

\label{sec:ablation}

To comprehensively validate our proposed method, we conducted ablation studies from two aspects: data strategy  and model architecture. The comparison results are detailed in the table.\ref{tab:transfer_analysis} and \ref{tab:ablation}.

\begin{table*}[t]
    \centering
    \small 
    \renewcommand{\arraystretch}{1.1} 
    \setlength{\tabcolsep}{0pt} 

    \caption{\textbf{Cross-Dialect Transfer Analysis.} Comparison between single-dialect Ti-Audio variants and the unified multi-dialect Ti-Audio.}
    \label{tab:transfer_analysis}
    
    \begin{tabularx}{\textwidth}{@{\extracolsep{\fill}} l l cccc c @{}}
        \toprule
        \textbf{Task} & \textbf{Training Source} & \textbf{Amdo} & \textbf{Kham} & \textbf{Ü-Tsang} & \textbf{Avg} & \textbf{Metric} \\
        \midrule
        \multirow{4}{*}{\textbf{ST}} 
            & \textbf{Ti-Audio} & \textbf{20.59} & \textbf{22.11} & \textbf{23.45} & \textbf{22.05} & BLEU \\
            & Amdo-only  & \phantom{0}6.85 & \phantom{0}1.41 & \phantom{0}0.87 & \phantom{0}3.04 & BLEU \\
            & Kham-only  & \phantom{0}2.14 & \phantom{0}9.90 & \phantom{0}5.00 & \phantom{0}5.68 & BLEU \\
            & Ü-Tsang-only & \phantom{0}0.50 & \phantom{0}1.91 & \phantom{0}6.90 & \phantom{0}3.10 & BLEU \\
        
        \addlinespace[0.5em] 
        \midrule
        \addlinespace[0.5em]
        
        \multirow{4}{*}{\textbf{ASR}} 
            & \textbf{Ti-Audio} & \textbf{14.25 / \phantom{0}9.57} & \textbf{14.15 / \phantom{0}9.60} & \textbf{14.99 / 10.31} & \textbf{14.46 / \phantom{0}9.83} & WER/CER (\%) \\
            & Amdo-only  & 95.94 / 80.17 & 185.67 / 157.95 & 190.69 / 169.53 & 157.43 / 135.88 & WER/CER (\%) \\
            & Kham-only  & 97.94 / 74.21 & \phantom{0}56.20 / \phantom{0}39.75 & \phantom{0}80.35 / \phantom{0}58.98 & \phantom{0}78.16 / \phantom{0}57.65 & WER/CER (\%) \\
            & Ü-Tsang-only & 226.49 / 209.23 & 202.30 / 190.02 & 108.31 / \phantom{0}92.33 & 179.03 / 163.86 & WER/CER (\%) \\
        \bottomrule
    \end{tabularx}

    \vspace{1.5em} 

    \caption{\textbf{Ablation Study on key components.} Analysis of the impact of Temperature Sampling, Dynamic Selection mechanism, and the overall Adapter architecture.} 
    \label{tab:ablation}
    \begin{tabularx}{\textwidth}{@{\extracolsep{\fill}} l l cccc c @{}}
        \toprule
        \textbf{Task} & \textbf{Configuration} & \textbf{Amdo} & \textbf{Kham} & \textbf{Ü-Tsang} & \textbf{Avg} & \textbf{Metric} \\
        \midrule
        \multirow{2}{*}{\textbf{ST}} 
            & \textbf{Ti-Audio}       & \textbf{20.59} & \textbf{22.11} & \textbf{23.45} & \textbf{22.05} & BLEU \\
            & -- w/o Adapter (Linear) & 15.21 & 15.23 & 15.67 & 15.37 & BLEU \\
        
        \addlinespace[0.5em]
        \midrule
        \addlinespace[0.5em]
        
        \multirow{2}{*}{\textbf{ASR}} 
            & \textbf{Ti-Audio}       & \textbf{14.25 / \phantom{0}9.57} & \textbf{14.15 / \phantom{0}9.60} & \textbf{14.99 / 10.31} & \textbf{14.46 / \phantom{0}9.83} & WER/CER (\%) \\
            & -- w/o Adapter (Linear) & 26.86 / 16.44 & 27.60 / 17.28 & 28.07 / 17.35 & 27.51 / 17.02 & WER/CER (\%) \\
        \bottomrule
    \end{tabularx}
\end{table*}

\subsection{Cross-Dialect Data Augmentation}
Recent research indicates that leveraging dialectal affinity significantly improves text-based tasks \cite{wang2023comprehensive}. 
Whether this collaborative mechanism can be applied to the voice domain to alleviate the problem of data imbalance remains to be explored.

To evaluate the effect of cross-dialect cooperation, we construct three controlled single-dialect baselines: Amdo-only, Kham-only, and Ü-Tsang-only. All these baselines use the same architecture and training objective as Ti-Audio, but are trained only on the corresponding dialect data without multi-dialect mixing or temperature-aware balancing. This setting allows us to isolate the contribution of unified multi-dialect training.

\begin{minipage}[t]{0.48\textwidth}
\vspace{0pt} 
\raggedright 

As shown in Figure~\ref{fig:dialect_lda}, the LDA feature projection visually reveals the relationships and spatial distribution among the three Tibetan language dialects. The Kham dialect is located in a transitional position in spatial distribution, with its center point situated between the Tibetan/Lhasa dialect and the Amdo dialect.

Quantitative results show that the Kham dialect is acoustically most closely related to Tibetan dialects (2.69),while also maintaining close ties with Amdo dialects (3.44). This distribution pattern acoustically confirms the transitional role of the Kham dialect within the Tibetan dialect chain. Based on this acoustic evidence, Table~\ref{tab:transfer_analysis} further demonstrates the effectiveness of our method for low-resource languages.
\end{minipage}
\hfill
\begin{minipage}[t]{0.48\textwidth}
\vspace{0pt} 
\centering

\includegraphics[width=\linewidth]{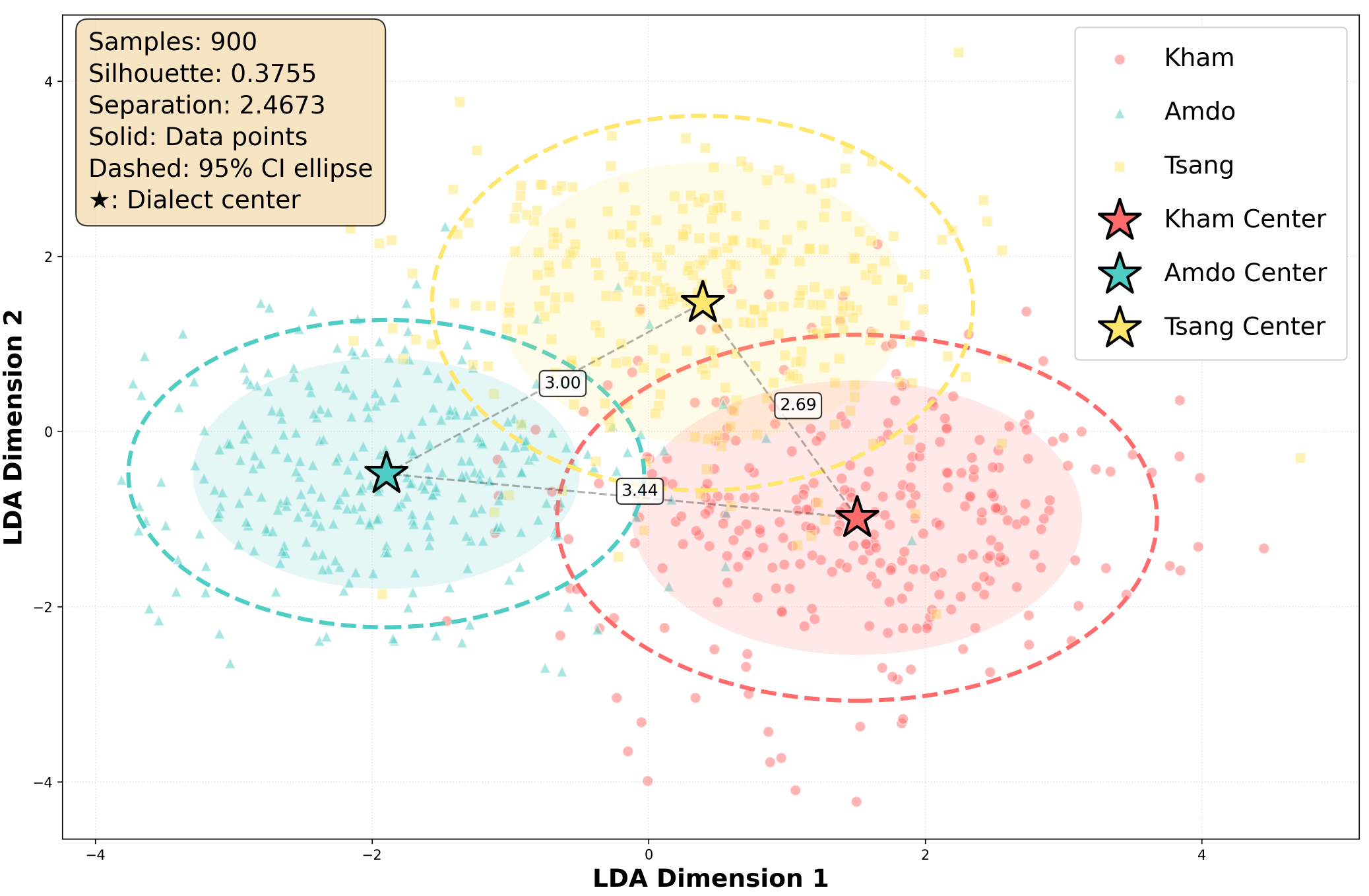}

\captionof{figure}{\textbf{Tibetan Three-Dialect Distribution}}
\label{fig:dialect_lda}
\end{minipage}

\subsubsection{Empirical Validation of the Linguistic Pivot Effect}

The experimental results show that the Kham model outperforms other dialects in both automatic speech recognition (ASR) and speech translation (ST) tasks. Particularly noteworthy is that the model trained solely with the Kham language demonstrates significantly stronger zero-sample transfer capabilities for the Amdo dialect and Tibet language compared to the reciprocal transfer capabilities between the latter two. This indicates that the Kham dialect plays a unique role as a language hub. 
 This collaborative effect centered on the Kham language helps the model achieve a broader generalization ability. Ti-Audio has learned more cross-features through cross-dialect training, confirming that this is an effective data augmentation method.

\subsection{Impact of the Adapter Architecture}

In order to rigorously verify the necessity of the architecture framework proposed in this study, we conducted an ablation study. In the experiment, the proposed adapter was replaced by the linear projector of the baseline model.

\subsubsection{Mechanism of the Linear Projector} 
We adopted a minimalist alignment module as the comparison baseline, which eliminated the complex aggregation mechanism in the original adapter. Under this configuration, the overall architecture simplifies to a single affine transformation layer, defined as $W \in \mathbb{R}^{d_{audio} \times d_{llm}}$. This layer linearly projects the hidden states extracted by the audio encoder (mHuBERT) into the semantic embedding space of the large language model (LLM). The post-projected acoustic features are then concatenated with the text embeddings from the input layer. This configuration (common in basic multimodal research) is designed to evaluate the effectiveness of speech translation and recognition tasks in such scenarios.

\subsubsection{Performance Analysis}

The results in the bottom part of Table \ref{tab:ablation} indicate that a simple linear mapping is completely unable to bridge the modal differences in this scenario. Replacing the adapter with a linear baseline leads to a significant drop in performance. The average ST BLEU score decreased from 22.05 to 15.37, and the ASR WER increased from 14.46\% to 27.51\%, almost doubling. This highlights the crucial role of the proposed adapter in aligning fine-grained acoustic details with semantic representations. At the same time, it ensures robustness in translation and recognition tasks.

\subsubsection{Adapter Efficiency Analysis}

\begin{figure}[t]
    \centering
    \begin{subfigure}{0.36\linewidth}
        \centering
        \includegraphics[width=\linewidth]{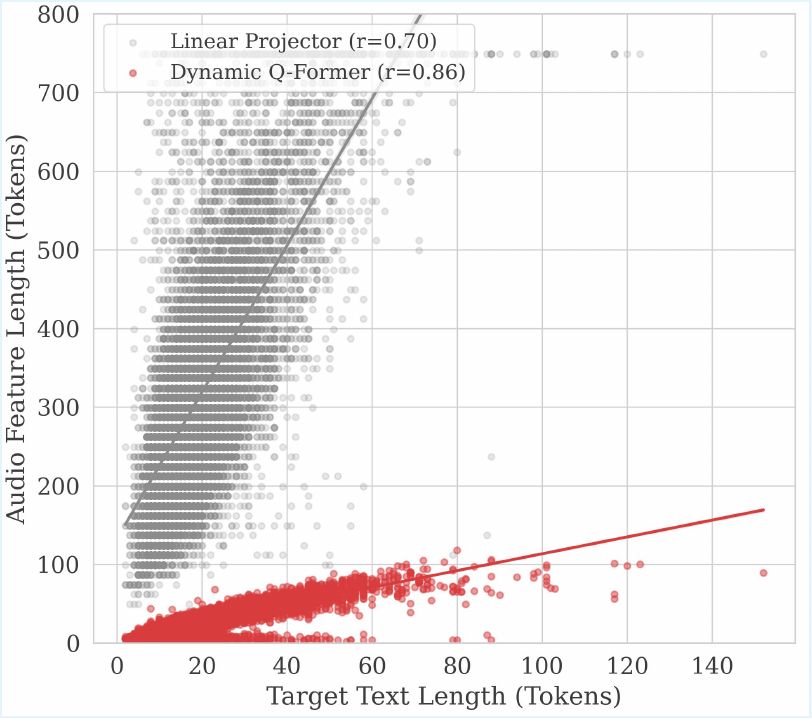}
        \caption{Length Correlation} 
    \end{subfigure}
    \quad 
    \begin{subfigure}{0.38\linewidth}
        \centering
        \includegraphics[width=\linewidth]{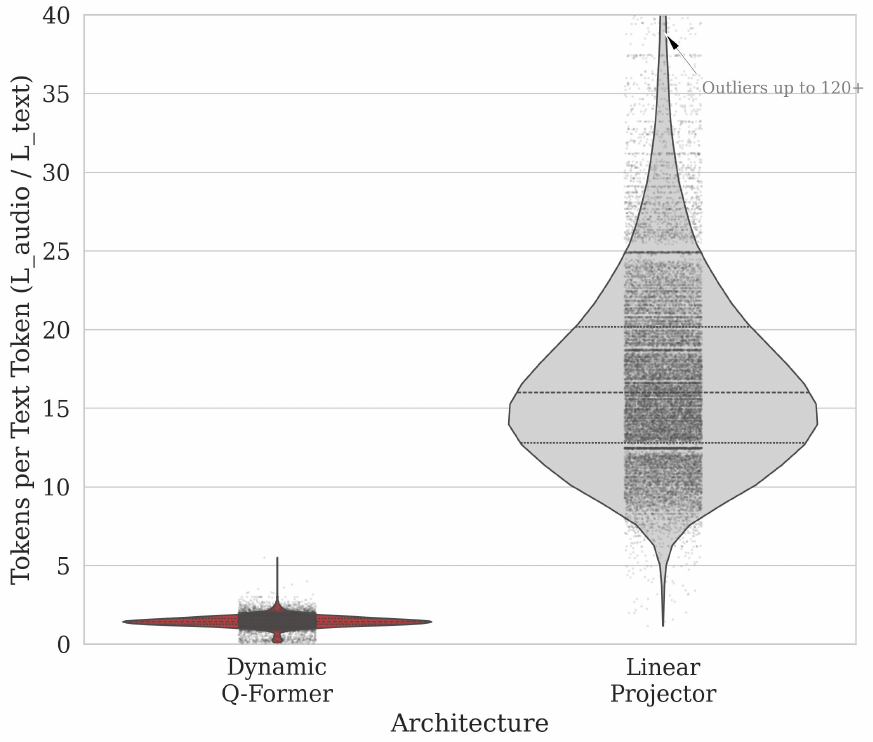}
        \caption{Token Efficiency} 
    \end{subfigure}
    \caption{Adapter Efficiency Analysis}
    \label{fig:efficiency}
\end{figure}

To evaluate the architecture efficiency, we analyzed the token expansion factor by calculating the ratio of audio tokens to text tokens ($L_{\text{audio}}/L_{\text{text}}$). Figure \ref{fig:efficiency}(a) shows that the Dynamic Q-Former achieved a higher length correlation ($r=0.86$ vs. $0.70$). This indicates a tighter modal alignment. Additionally, Figure \ref{fig:efficiency}(b) indicates that our adapter effectively extracts redundant acoustic signals into a compact ratio of approximately 2.0, while the linear baseline remains sparse and redundant (with an average of 17.5). Ultimately, these results reflect that the Dynamic Q-Former outperforms the linear projector in extracting compact semantic features. It significantly enhances cross-modal bridging while reducing sequence redundancy by nearly 90\%.

\section{Conclusion}

In this paper, we propose Ti-Audio, a Tibetan speech-language large model for low-resource and multi-dialect speech understanding. By introducing a Dynamic Q-Former Adapter and a temperature-aware sampling strategy, Ti-Audio achieves advanced performance in ASR and ST tasks. The Dynamic Q-Former Adapter effectively bridges acoustic representations and large language models, while the temperature-aware sampling strategy alleviates dialectal data imbalance and improves robustness to phonetic variations. Experimental results demonstrate that Ti-Audio outperforms most baselines and shows strong generalization across heterogeneous Tibetan dialects. These findings indicate that combining cross-modal adaptation with dialect-balanced training provides an effective solution for Tibetan speech understanding and future low-resource speech large models.

\bibliographystyle{unsrt}  
\bibliography{references}

\end{document}